\documentclass[dvipsnames,letterpaper, 10 pt, journal]{ieeeconf}

\IEEEoverridecommandlockouts	

\usepackage{cite}
\usepackage{amsmath,amssymb,amsfonts}
\usepackage{algorithmic}
\usepackage{textcomp}
\usepackage{graphicx,color}
\usepackage{mathrsfs}
\usepackage{tikz}
\usepackage[vlined,ruled]{algorithm2e}
\usepackage{subfigure}
\usepackage{url}
\usepackage{color}
\usepackage{dsfont}
\usepackage{bbm}
\usepackage{booktabs}
\usepackage{array}
\usepackage{yfonts}
\usepackage[normalem]{ulem}
\usepackage{arydshln,leftidx,mathtools}
\usepackage{multicol}
\usepackage{amsmath}
\usepackage{stfloats}
\usepackage{comment}
\usepackage{subfigure}
\usepackage{manyfoot}
\usepackage{listings}

\definecolor{myBlue}{RGB}{49,130,189}

\newcommand{\map}[3]{#1: #2 \rightarrow #3}

\usepackage{tabularx}
\usepackage{multirow}
\usepackage{makecell}

\newcommand\aamsout{\bgroup\markoverwith{\textcolor{violet}{\rule[0.5ex]{2pt}{1pt}}}\ULon}

\DeclareSymbolFont{bbold}{U}{bbold}{m}{n}
\DeclareSymbolFontAlphabet{\mathbbold}{bbold}

\newcommand\oprocendsymbol{\hbox{$\square$}}
\newcommand\oprocend{\relax\ifmmode\else\unskip\hfill\fi\oprocendsymbol}

\renewcommand{\theenumi}{(\roman{enumi})}

\newcommand*{\QEDA}{\hfill\ensuremath{\blacksquare}}%

\DeclareNewFootnote{R}[roman]

\graphicspath{{figs/}}

\makeatletter
\let\NAT@parse\undefined
\makeatother
\usepackage[colorlinks,urlcolor=blue, linkcolor=blue, citecolor=blue]{hyperref}

\begin{document}

\title{\LARGE \bf Data-Efficient Physics-Informed Learning to Model Synchro-Waveform Dynamics of Grid-Integrated Inverter-Based Resources}


\author{ Shivanshu~Tripathi, Hossein~Mohsenzadeh~Yazdi, Maziar~Raissi, and Hamed~Mohsenian-Rad\vspace{-0.5cm}
 \thanks{
 S.~Tripathi, H.~M.~Yazdi and H.~Mohsenian-Rad are with the Department of Electrical and Computer Engineering, and M.~Raissi is with the Department of Mathematics, University of California, Riverside, CA, USA. E-mails: \{strip008, hmohs003, hamedrad, maziarr\}@ucr.edu.
%
    %
    The corresponding authors are M.~Raissi and H.~Mohsenian-Rad.}}

\maketitle
\pagestyle{empty}
\thispagestyle{empty}

\begin{abstract}
Inverter-based resources (IBRs) exhibit fast transient dynamics during network disturbances, which often cannot be properly captured by phasor and SCADA measurements. This shortcoming has recently been addressed with the advent of waveform measurement units (WMUs), which provide high-resolution, time-synchronized raw voltage and current waveform samples from multiple locations in the power system. However, transient model learning based on synchro-waveform measurements remains constrained by the scarcity of network disturbances and the complexity of the underlying nonlinear dynamics of IBRs. We propose to address these problems by developing a data-efficient physics-informed machine learning (PIML) framework for synchro-waveform analytics that estimates the IBR terminal current response from only a few network disturbance signatures. Here, the physics of the electrical circuits are used to compensate for limited data availability by constraining the learning process through known circuit relationships. Two cases are considered, with known and unknown circuit parameters. In the latter case, the framework jointly learns the transient dynamics of the IBRs and the parameters of the electrical circuit. Case studies using WMU disturbance data across multiple sampling rates shows consistently lower current estimation error with substantially fewer training events than a purely data-driven baseline.

\vspace{0.2cm}

\textbf{\emph{Keywords:} Synchro-Waveform measurements, inverter-based resources, transient dynamics, data-driven model, network disturbances, physics-informed learning, data efficiency.}

\end{abstract}


\vspace{-0.2cm}

\section{Introduction}\label{sec: introduction}

 
The increasing penetration of inverter-based resources (IBRs) is rapidly transforming modern power systems from synchronous generation to power-electronics-based, asynchronous generation. IBRs respond to grid disturbances through fast, control-driven dynamics that often include significant sub-cycle transients. Recent reports from the North American Electric Reliability Corporation (NERC) have documented several instances of unexpected IBR behavior during grid disturbances, underscoring the need for improved modeling of the transient dynamics of IBRs \cite{NERC_report_2}.

Motivated by this need, high-resolution, GPS time-synchronized waveform measurements from waveform measurement units (WMUs) have been deployed at IBRs to model their transient dynamics at the level of raw voltage and current waveform samples. WMUs are, in essence, an extension of traditional power quality (PQ) meters and digital fault recorders (DFRs) that provide waveform samples with GPS time synchronization, either in an event-triggered format or, more recently, through continuous streaming \cite{synchro_waveform_window}. These measurements enable direct modeling of fast sub-cycle IBR dynamics using data from multiple locations \cite{Hossein_letter,subcycle_overvoltage}.


\begin{figure}[t]
  \centering
  \includegraphics[width=1\columnwidth,trim={0cm 0cm 0cm
    0cm},clip]{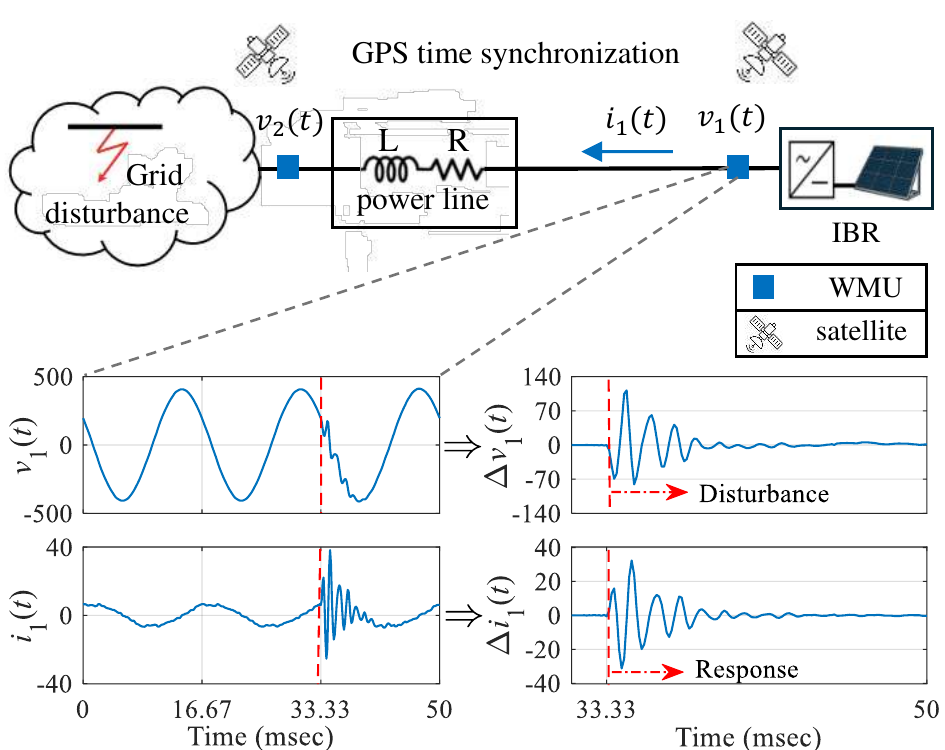} \vspace{-0.7cm}
  \caption{Grid-connected IBR monitored by GPS time-synchronized WMUs, showing the transient voltages $v_1(t),v_2(t)$ and current response $i_1(t)$ following grid disturbances across line with resistance~$R$ and inductance~$L$.\vspace{-0.3cm}}
    \label{fig: waveform}
\end{figure}

Despite advancements in measurement capabilities, learning accurate IBR transient dynamic models from a \emph{small number} of WMU-recorded disturbances remains challenging. On the one hand, \emph{physics-only} approaches can be computationally burdensome or sensitive to uncertainty at sub-cycle time scales. On the other hand, \emph{data-only} models require large labeled disturbance datasets that are rarely available.

To bridge this gap, this paper proposes a novel \emph{physics-informed machine learning} (PIML) approach. The basic idea is to leverage additional \emph{grid-side} waveform measurements that are already available but have so far not been used for this type of analysis. These measurements are combined with IBR-site waveform measurements, enabling the incorporation of circuit equations into a PIML model framework to constrain the learned dynamics. See, e.g., the setup in Fig. \ref{fig: waveform}. The details of this setup are explained in Section II.

\vspace{0.1cm}

\textbf{\emph{Related Work – IBR Modeling}:}
The existing literature on IBR modeling can be categorized into two classes, namely \emph{physics-based modeling} and \emph{data-driven modeling}. Physics-based models can capture electromagnetic and control dynamics \cite{physics_1,physics_2}. However, they become computationally expensive when fast timescales, such as sub-cycle transients, are involved. Their accuracy may also degrade when the underlying physics is complex \cite{cai2016modelling}.
This shortcoming can be addressed using data-driven models. Both phasor measurements \cite{palberz} and waveform measurements \cite{Fatemeh_letter,Hossein-Journal, synchro_waveform_mladen} have been used for this purpose. However, data-driven models often require large volumes of labeled disturbance data. Built upon WMU data, the method in \cite{Fatemeh_letter} uses regression models, while the methods in \cite{Hossein-Journal} use Long Short-Term Memory (LSTM) models combined with pre-disturbance operating conditions. Nevertheless, both \cite{Fatemeh_letter} and \cite{Hossein-Journal} require access to waveform data from a large number of disturbances, which may require waiting several months to collect sufficient labeled data.

\vspace{0.1cm}

\textbf{\emph{Related Work - PIML in Power Systems}:} We propose a novel PIML framework to address the above real-world challenges. PIML is a promising framework that integrates physical laws into the learning process \cite{MR-PP-GEK:19}. In power systems, PIML has been used in a range of problems, including optimal power flow \cite{OPF-PINN}, anomaly detection \cite{Data_Anomaly_PINN}, and state estimation \cite{DSE_PINN,RPSSE_PINN,Line_par_PINN,GSM-AV-SC:20}. To the best of our knowledge, this is the first study to use PIML for synchro-waveform analytics.



\vspace{0.1cm}

\textbf{\emph{Technical Contributions}:} 
\emph{First}, we propose a novel PIML approach that improves data efficiency in waveform-based IBR model learning. The proposed approach significantly enhances accuracy in learning the dynamic response of an IBR to transient disturbances, while using substantially fewer labeled instances than a purely data-driven baseline. \emph{Second}, we simultaneously estimate the \emph{unknown} power line parameters within the same PIML framework. \emph{Third}, we validate our methods through a case study across \emph{different sampling rates}. The results demonstrate consistently lower estimation error and robust performance even at low sampling rates.

 \section{Problem formulation}\label{sec: formulation}


Again consider the setup in Fig. \ref{fig: waveform} with two WMUs. One WMU is installed at the IBR to measure its terminal voltage $v_1(t)$ and current $i_1(t)$. Another WMU is installed at the grid-side of the power line to measure voltage $v_2(t)$. These measurements correspond to a disturbance that produces a short-duration voltage oscillation at the IBR terminal, which in turn excites dynamic response in the inverter current.

Let $x(t)\in\{v_1(t), i_1(t), v_2(t)\}$ denote a set of measured waveforms during a network disturbance. Rather than working directly with the raw waveforms, we use the first-order differences of the waveform samples, i.e., the following differential waveforms \cite[p.~151]{Hamed_book}:
$\Delta x(t) \triangleq x(t)-x(t-N \Delta t)$, at any $t\ge t_0$,
where $\Delta t$ is the sampling interval, $N$ is the number of samples per cycle, and $t_0$ is the onset time of the disturbance. Focusing on $\Delta x(t)$ rather than $x(t)$ ensures that we examine only the disturbance signatures in all waveforms, and not the pre-disturbance steady-state waveforms. This highlights the transient content associated with both the disturbance and the IBR response.


In this work, we aim to learn the relationship between the differential disturbance and the differential IBR response:
\begin{equation}\label{eq:IBR}
\Delta i_1(t)=f\big(\Delta v_1(t)\big), \quad t\ge t_0,
\end{equation}
where $\map{f}{\mathbb{R}}{\mathbb{R}}$ is an unknown nonlinear function
with input $\Delta v_1(t)\in\mathbb{R}$. We use the fact that the differential waveforms satisfy the grid-side circuit relationship:
\begin{align}\label{eq:kirchhoff_cont}
\Delta v_2(t)=\Delta v_1(t)-R\,\Delta i_1(t)-L\,\frac{d \Delta i_1}{dt}, \quad t\geq t_0,
\end{align} 
where $R$ and $L$ are line resistance and inductance. 

We collect time-synchronized measurements from $p$ disturbance events, indexed by $k\in\{1,\ldots,p\}$. Let $\ell\in\{0,\ldots,n-1\}$ denote the sample index within an event, and define the sampling instants, $t_\ell=t_0+\ell\Delta t$. The discrete-time measurement corresponding to a continuous-time signal is given as, $x[\ell]\triangleq x(t_\ell)$. The corresponding differential samples are given as
$\Delta x[\ell]=x[\ell]-x[\ell-N]$, for $\ell>1$.
With $n$ samples per event and 
$p$ events in total, the raw dataset $x[\ell]$ collected by WMUs is converted to differential measurements $\Delta x[\ell]$, which is represented as:
\begin{align}
\mathcal{D} := \big\{ \big\{ \Delta v_{1,k}[\ell],\Delta v_{2,k}[\ell],\Delta i_{1,k}[\ell], \ell \big\}_{\ell=0}^{n-1} \big\}_{k=1}^p.
\label{eq:dataset}
\end{align}

We consider two variations of the above problem: 
\begin{enumerate}
    \item Circuit parameters $R$ and $L$ are \emph{known}. We only need to learn the IBR current response map $f(\cdot)$.

    \item Circuit parameters $R$ and $L$ are \emph{unknown}. We need to simultaneously estimate map $f(\cdot)$ as well as line parameters $R$ and $L$ to have an accurate IBR model. 
\end{enumerate}



\color{black}

 \section{Estimating the IBR Terminal Current map}\label{eq: IBR_map}

 
In this section, we present our approach for learning the IBR response map $f(\cdot)$ in~\eqref{eq:IBR} under both settings (i) the (ii) that we previously outlined at the end of Section \ref{sec: formulation}. We will later compare our PIML approach that incorporates both the known circuit equation \eqref{eq:kirchhoff_cont} and the differential measurements \eqref{eq:dataset}, against the data-driven baseline approach. We denote the resulting estimates of the terminal differential response by $\Delta \hat i_1^{\text{phy}}$ (physics-informed) and $\Delta \hat i_1^{\text{data}}$ (data-driven). 

We partition the $p$ disturbance events into disjoint training, validation, and test sets with index sets $\mathcal{P}_{\mathrm{train}},\mathcal{P}_{\mathrm{val}}$, and $\mathcal{P}_{\mathrm{test}}$, where $[\mathcal{P}_{\mathrm{train}},\mathcal{P}_{\mathrm{val}},\mathcal{P}_{\mathrm{test}}]\subseteq\{1,\ldots,p\}$. Performance is evaluated using the mean-squared error (MSE), averaged over samples on all the test events:
\begin{align}\label{eq:test}
\mathcal{L}_{\text{test}}(\Delta \hat i_1)
:= \mathbb{E}_{k\sim\mathcal{P}_{\mathrm{test}}}
\left[\frac{1}{n}\sum_{\ell=0}^{n-1}
\bigl\|\Delta\hat i_{1}[\ell]- \Delta i_{1,k}[\ell]\bigr\|_2^2\right],
\end{align}
where $\Delta\hat i_1\in\{\Delta\hat i_1^{\mathrm{data}},\,\Delta\hat i_1^{\mathrm{phy}}\}$. We next detail the learning formulation for each of the two parameter settings:
\subsection{$R$ and $L$ are known}\label{known}
 For the physics-informed model, we estimate the differential IBR response using two neural networks of comparable size. We evaluate the proposed physics-informed approach against a purely data-driven model trained using the same training, validation and testing data. Next, we describe the procedure used to estimate the differential IBR response for both: data-only and physics-informed model.

\emph{Data-only model:} As a baseline, we train a single neural network on the IBR training set $\{(\Delta v_{1,k}[\ell],\,\Delta i_{1,k}[\ell],\, \ell)\}_{\ell=0}^{n-1}$ for all $k\in\mathcal{P}_{\mathrm{train}}$. Given the input pair $\{(\Delta v_{1,k}[t_\ell],t_\ell)\}$, the network predicts the terminal differential current: 
 \begin{align}
 \Delta \hat i_1^{\text{data}}[\ell]=\text{NN}_\text{data\_only}(\Delta v_{1,k}[\ell],\ell).
 \end{align}
The neural network parameters are learned by minimizing the empirical MSE over the training events as
 \begin{align}\label{eq: data-only}
\mathcal{L}_{\text{data}}
:= \mathbb{E}_{k\sim\mathcal{P}_{\mathrm{train}}}
\left[\frac{1}{n}\sum_{\ell=0}^{n-1}
\bigl\|\Delta \hat i_{1}^{\text{data}}[\ell]- \Delta i_{1,k}[\ell]\bigr\|_2^2\right].
\end{align}

\emph{Physics-informed model:} This model predicts the terminal IBRs response using two neural networks of comparable size. The first neural network is trained using the training set \eqref{eq:dataset}. 
The predicted current using the first neural network for the physics-informed model is given by:
 \begin{align}
 \Delta\hat i_1^{\text{phy}}[\ell] = \text{NN}_{\text{data}}(\Delta v_{1,k}[\ell],\ell),
 \end{align}

 The second neural network enforces equation \eqref{eq:kirchhoff_cont} as:
 \begin{align}\label{eq: pinn-constraints}
 \text{NN}_{\mathrm{phy}}(&\Delta v_{1,k}[\ell],\Delta v_{2,k}[\ell],\Delta \hat i_1^{\text{phy}}[\ell])= \Delta v_{2,k}[\ell]-\nonumber\\&\Delta v_{1,k}[\ell]+R\,\Delta \hat i^{\mathrm{phy}}_1[\ell]+L\,\frac{d }{dt}\Delta \hat i_1^{\mathrm{phy}}[\ell]=0,
 \end{align}
 where
  \begin{align}
 \frac{d }{dt}\Delta \hat i_1^{\text{phy}}[\ell]=\frac{\Delta \hat i^{\text{phy}}_1[\ell]-\Delta \hat i^{\text{phy}}_1[{\ell-1}]}{\Delta t}.
 \end{align} 
 We train the parameters of both the neural networks by minimizing the MSE over training data as:
\begin{align}
\mathcal{L}_{\text{Phy}}&:=\mathbb{E}_{k\sim\mathcal{P}_{\mathrm{train}}}
\bigg[\frac{1}{n}\sum_{\ell=0}^{n-1}
\bigl\|\Delta \hat i_{1}^{\text{phy}}[\ell]- \Delta i_{1,k}[\ell]\bigr\|_2^2\nonumber\\&+\lambda\bigl\| v_2[\ell]-v_1[\ell]+R\,\hat i^{\mathrm{phy}}_1[\ell]+L\,\frac{d }{dt}\hat i_1^{\mathrm{phy}}[\ell]  \bigl\|^2_2 \bigg],
\end{align}  
 where $\lambda$ is a hyperparameter chosen using validation dataset.

\subsection{$R$ and $L$ are Unknown}\label{unknown}
When the line parameters are unavailable, we treat $R$ and $L$ as learnable scalars and estimate them jointly with the IBR current response. We parameterize
$
R=\phi_R(\theta_R),\;
L=\phi_L(\theta_L),
$
where $\theta_R,\theta_L$ are trainable parameters
, and $\phi_R,\phi_L$ enforce feasibility ( $\phi(\cdot)=\mathrm{softplus}(\cdot)$ to ensure $R,L\ge 0$). We optimize them together with the neural-network weights via gradient-based training. The current predictor \(\text{NN}_{\text{data}}\) is identical to the known-parameter case in Section~\ref{known}. The only modification is that the circuit residual (used in the physics regularization term) is evaluated with the learned values of \((R,L)\). Thus, gradient-based training simultaneously estimate the differential IBR response and identifies the line parameters.
In this formulation, the physics term regularizes the learning problem and simultaneously identifies the effective line parameters from the data \eqref{eq:dataset}.

\section{Case Studies}
We consider the grid-connected IBR circuit monitored GPS time-synchronized WMUs in Fig.~\ref{fig: waveform},
with line parameters $R=10~\Omega$, and $L=0.2~mH$. 
We collect dataset in \eqref{eq:dataset} with $80$ events\footnote{{Open source code repository is made public at: \url{https://github.com/shivanshutripath/Data-Efficient-PIML-for-IBR}}}. 
We fix $10$ events for validation and $20$ events used for testing, and vary the number of training events in $\{3,5,10,20,30,40,50\}$. The validation and test sets are excluded from training and are used for selecting the hyper-parameter $\lambda$ and for performance
evaluation, respectively. 

We study both cases: (i) and (ii) that we defined in Section II, i.e., with known and with unknown $R$ and $L$. In each case, 
we evaluate multiple sampling resolutions to assess whether the
physics-informed estimator outperforms the data-only baseline; in the unknown-parameter case, we additionally estimate $R$ and $L$. Specifically, we consider $128$, $64$, and $32$ samples per $60$~Hz grid cycle, corresponding to 
\begin{align*}
\Delta t \in \left\{ \frac{1}{128\times 60},\ \frac{1}{64\times 60},\ \frac{1}{32\times 60} \right\}\ \text{s}.
\end{align*}

\begin{table}[t]
\centering
\caption{Performance comparison of data-only model and physics-informed model with known $R$ and $L$,  and Sampling Rate of $(128\times 60)$ Hz.}
\begin{tabular}{c c c c c}
\toprule
\textbf{TrainEv} & 
\textbf{$\mathcal{L}_{\text{test}}(\Delta\hat i_1^{\text{data}})$} & 
\textbf{$\mathcal{L}_{\text{test}}(\Delta\hat i_1^{\text{phy}})$} & 
\textbf{Imp. (\%)} & 
\textbf{$\lambda$}  \\
\midrule
3  & $22.11$ & $9.30$ & $+57.94$ & $0.1$ \\
5  & $19.06$ & $8.85$ & $+53.56$ & $0.001$\\
10 & $14.56$ & $8.87$ & $+39.03$ & $3$ \\
20 & $10.83$ & $8.85$ & $+18.33$  & $1$ \\
30 & $9.89$ & $8.08$ & $+18.22$ & $0.3$ \\
40 & $8.69$ & $8.23$ & $+5.24$ & $0.3$\\
50 & $9.56$ & $8.00$ & $+16.29$  & $0.1$\\
\bottomrule
\end{tabular}
\label{tab:nn_vs_phys_128}
\end{table}

\subsection{Case I: Known $R$ and $L$}\label{subsec: knownRL}
In this setting, the line parameters $(R,L)$ are assumed known and we compare the \emph{data-only} and \emph{physics-informed} estimators described in Section~\ref{known}. The data-only model uses a single neural network, whereas the physics-informed model uses two; for a fair comparison, all networks share the same architecture: two hidden layers with $32$ neurons per layer and $\tanh$ activations.

\begin{figure*}[!t]
  \centering
  \includegraphics[width=2\columnwidth,trim={0cm 0cm 0cm
    0cm},clip]{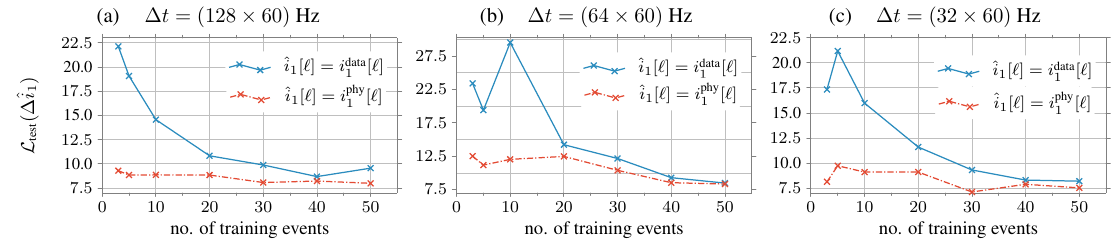}
  \caption{
  The MSE for the testing data using data-only model (solid blue) and the PHIL model (dashed red) for \emph{known} circuit parameters; see Section~\ref{subsec:
    knownRL}. The PIML model achieves lower prediction error than the data-only baseline. (a)-(c) correspond to sampling rates of 128, 64, and 32 samples per cycle.}
    \label{fig: all_fig_1}
\end{figure*}

Tables~\ref{tab:nn_vs_phys_128}-\ref{tab:nn_vs_phys_32} report the performance of both models for sampling rates of $(128\times 60)$~Hz, $(64\times 60)$~Hz, and $(32\times 60)$~Hz, respectively. Fig.~\ref{fig: all_fig_1} summarizes the average MSE of the estimated differential IBR current response across training-set sizes for each sampling rate. Across all resolutions, the physics-informed model achieves consistently lower MSE than the data-only baseline. Moreover, Table~\ref{tab:data_eff} quantifies the data-efficiency gain: to attain comparable error levels, the physics-informed approach requires approximately $7$--$9$ times fewer training events than the data-only model.

\begin{table}[t]
\centering
\caption{Performance comparison of data-only model and physics-informed model with known $R$ and $L$,  and Sampling Rate of $(64\times 60)$ Hz.}
\begin{tabular}{c c c c c}
\toprule
\textbf{TrainEv} & 
\textbf{$\mathcal{L}_{\text{test}}(\Delta\hat i_1^{\text{data}})$} & 
\textbf{$\mathcal{L}_{\text{test}}(\Delta\hat i_1^{\text{phy}})$} & 
\textbf{Imp. (\%)} & 
\textbf{$\lambda$}  \\
\midrule
3  & $23.38$ & $12.50$ & $+47.65$ & $0.1$ \\
5  & $19.38$ & $11.17$ & $+42.35$ & $1$\\
10 & $29.84$ & $12.03$ & $+59.67$ & $0.0003$ \\
20 & $14.22$ & $12.47$ & $+12.30$  & $0.003$ \\
30 & $12.16$ & $10.39$ & $+14.58$ & $0.001$ \\
40 & $9.29$ & $8.54$ & $+8.07$ & $0.03$\\
50 & $8.48$ & $8.34$ & $+1.68$  & $0.1$\\
\bottomrule
\end{tabular}
\label{tab:nn_vs_phys_64}
\end{table}

\begin{table}[t]
\centering
\caption{Performance comparison of data-only model and physics-informed model with known $R$ and $L$,  and Sampling Rate of $(32\times 60)$ Hz.}
\begin{tabular}{c c c c c}
\toprule
\textbf{TrainEv} & 
\textbf{$\mathcal{L}_{\text{test}}(\Delta\hat i_1^{\text{data}})$} & 
\textbf{$\mathcal{L}_{\text{test}}(\Delta\hat i_1^{\text{phy}})$} & 
\textbf{Imp. (\%)} & 
\textbf{$\lambda$}  \\
\midrule
3  & $17.35$ & $8.12$ & $+53.20$ & $0.1$ \\
5  & $21.18$ & $9.72$ & $+55.52$ & $0.003$\\
10 & $15.98$ & $9.09$ & $+36.75$ & $0.1$ \\
20 & $11.59$ & $9.09$ & $+21.53$  & $0.1$ \\
30 & $9.30$ & $7.10$ & $+23.69$ & $1$ \\
40 & $8.29$ & $7.87$ & $+5.40$ & $0.3$\\
50 & $8.18$ & $7.51$ & $+8.14$  & $0.01$\\
\bottomrule
\end{tabular}
\label{tab:nn_vs_phys_32}
\end{table}

\begin{table}[t]
\centering
\caption{Data efficiency of the physics informed model compared to the data-only model (baseline).}
\begin{tabular}{c c | c c}
\toprule
\multicolumn{2}{c|}{\textbf{Known $R,L$}} & \multicolumn{2}{c}{\textbf{Unknown $R,L$}} \\
\cmidrule(lr){1-2}\cmidrule(lr){3-4}
\textbf{Sampling rate} &
\textbf{data-efficiency} &
\textbf{Sampling rate} &
\textbf{data-efficiency} \\
\midrule
128 & $\approx 7\times$ & 128 & $\approx 2.34\times$ \\
64  & $\approx 9\times$ & 64  & $\approx 5.67\times$ \\
32  & $\approx 7\times$ & 32  & $\approx 2.34\times$ \\
\bottomrule
\end{tabular}
\label{tab:data_eff}
\end{table}

\begin{figure*}[!t]
  \centering
  \includegraphics[width=2\columnwidth,trim={0cm 0cm 0cm
    0cm},clip]{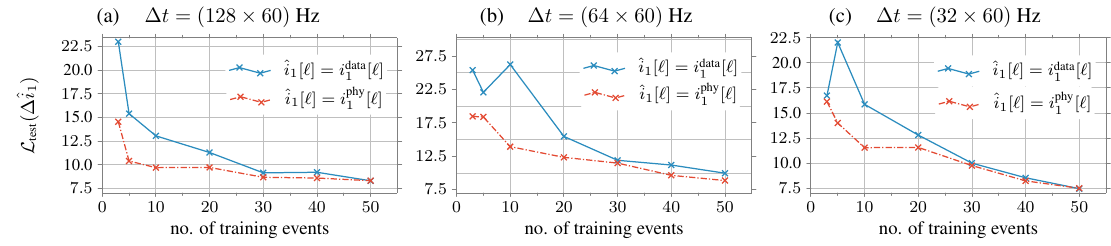}
  \caption{
  The MSE for the testing data using data-only model (solid blue) and the PIML model (dashed red) for \emph{unknown} circuit parameters; see Section~\ref{subsec:
    unknownRL}. The PIML model achieves lower prediction error than the data-only baseline. (a)-(c) correspond to sampling rates of 128, 64, and 32 samples per cycle.}
    \label{fig: all_fig_5}
\end{figure*}

\begin{figure}[!t]
  \centering
  \includegraphics[width=1\columnwidth,trim={0cm 0cm 0cm
    0cm},clip]{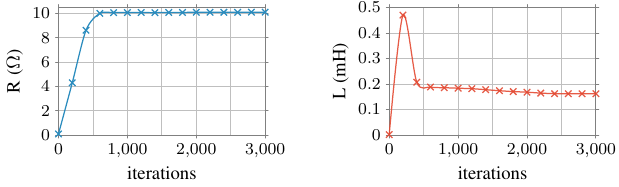}
  \caption{
  The learned resistance $R$
   and inductance $L$ over the training iterations 
   with 20 training events. Both 
   parameters are updated through gradient-based 
   optimization and gradually converge to true parameters.}
   \vspace{-0.4cm}
    \label{fig: example_3_1}
\end{figure}

\subsection{Case II: Unknown $R$ and $L$}\label{subsec: unknownRL}
We adopt the same experimental setup as in Section~\ref{subsec: knownRL}, but now treat $(R,L)$ as unknown and estimate them jointly with the differential IBR current response. This enables simultaneous recovery of $\Delta \hat i_1(t)$ and identification of $(R,L)$.

Tables~\ref{tab:nn_vs_phys_4}--\ref{tab:nn_vs_phys_6} report the MSE of differential current response estimation and the corresponding estimates of $(R,L)$ across sampling rates. Fig.~\ref{fig: all_fig_5} summarizes the average MSE versus the number of training events, and Fig.~\ref{fig: example_3_1} shows the evolution of the learned $(R,L)$ during 20 training events. Overall, the PIML model has better performance and  higher data-efficiency than the data-only baseline shown in Table~\ref{tab:data_eff}, while additionally identifying the circuit parameters. 

\begingroup
\setlength{\textfloatsep}{6pt plus 1pt minus 2pt} 
\setlength{\floatsep}{4pt plus 1pt minus 2pt}     
\setlength{\abovecaptionskip}{2pt}
\setlength{\belowcaptionskip}{2pt}
\begin{table*}[t]
\centering
\caption{Performance comparison of data-only model and physics-informed model with\\ unknown $R$ and $L$, and Sampling rate of $(128\times60)$ Hz}
\begin{tabular}{c c c c c c c}
\toprule
\textbf{TrainEv} &
\textbf{$\mathcal{L}_{\text{test}}(\Delta\hat i_1^{\text{data}})$} &
\textbf{$\mathcal{L}_{\text{test}}(\Delta\hat i_1^{\text{phy}})$} &
\textbf{Improvement (\%)} &
\textbf{Resistance (ohm)} &
\textbf{Inductance (mH)} &
\textbf{$\lambda$} \\
\midrule
3  & $22.99$ & $14.53$ & $+36.82$ & $9.898$ & $0.192$  & $0.0001$\\
5  & $15.39$ & $10.38$ & $+32.58$ & $9.698$ & $0.268$ & $1$\\
10 & $13.04$ & $9.68$ & $+25.70$ & $9.874$ & $0.178$ & $1$ \\
20 & $11.28$ & $9.70$ & $+13.93$ & $9.906$ & $0.169$ & $0.00001$\\
30 & $9.13$ & $8.66$ & $+5.17$ & $9.750$ & $0.220$ & $0.3$\\
40 & $9.29$ & $8.58$ & $+7.65$ & $9.017$ & $0.480$ & $1$\\
50 & $8.29$ & $8.30$ & $-0.06$ & $8.852$ & $0.548$ & $1$ \\
\bottomrule
\end{tabular}
\label{tab:nn_vs_phys_4}
\end{table*}

\begin{table*}[t]
\centering
\caption{Performance comparison of data-only model and physics-informed model with\\ unknown $R$ and $L$, and Sampling rate of $(64\times60)$ Hz}
\begin{tabular}{c c c c c c c}
\toprule
\textbf{TrainEv} &
\textbf{$\mathcal{L}_{\text{test}}(\Delta\hat i_1^{\text{data}})$} &
\textbf{$\mathcal{L}_{\text{test}}(\Delta\hat i_1^{\text{phy}})$} &
\textbf{Improvement (\%)} &
\textbf{Resistance (ohm)} &
\textbf{Inductance (mH)} &
\textbf{$\lambda$} \\
\midrule
3  & $25.37$ & $18.45$ & $+27.27$ & $9.736$ & $0.268$  & $0.001$\\
5  & $22.06$ & $18.37$ & $+16.74$ & $9.700$ & $0.295$ & $0.00001$\\
10 & $26.21$ & $13.93$ & $+46.87$ & $9.65$ & $0.302$ & $0.001$ \\
20 & $15.46$ & $12.33$ & $+20.25$ & $9.293$ & $0.403$ & $0.3$\\
30 & $11.89$ & $11.47$ & $+3.54$ & $9.411$ & $0.352$ & $0.03$\\
40 & $11.18$ & $9.64$ & $+13.74$ & $8.954$ & $0.494$ & $0.01$\\
50 & $9.96$ & $8.86$ & $+5.71$ & $8.857$ & $0.516$ & $0.0001$ \\
\bottomrule
\end{tabular}
\label{tab:nn_vs_phys_5}
\end{table*}

\begin{table*}[t]
\centering
\caption{Performance comparison of data-only model and physics-informed model with\\ unknown $R$ and $L$, and Sampling rate of $(32\times60)$ Hz}
\begin{tabular}{c c c c c c c}
\toprule
\textbf{TrainEv} &
\textbf{$\mathcal{L}_{\text{test}}(\Delta\hat i_1^{\text{data}})$} &
\textbf{$\mathcal{L}_{\text{test}}(\Delta\hat i_1^{\text{phy}})$} &
\textbf{Improvement (\%)} &
\textbf{Resistance (ohm)} &
\textbf{Inductance (mH)} &
\textbf{$\lambda$} \\
\midrule
3  & $16.73$ & $16.13$ & $+3.56$ & $8.600$ & $0.577$  & $0.3$\\
5  & $22.04$ & $13.99$ & $+36.54$ & $8.244$ & $0.674$ & $0.3$\\
10 & $15.86$ & $11.53$ & $+27.32$ & $6.867$ & $1.062$ & $0.3$ \\
20 & $12.79$ & $11.55$ & $+9.71$ & $6.246$ & $1.244$ & $0.001$\\
30 & $9.98$ & $9.74$ & $+2.50$ & $6.363$ & $1.197$ & $0.00001$\\
40 & $8.52$ & $8.21$ & $+3.58$ & $6.576$ & $1.127$ & $0.001$\\
50 & $7.43$ & $7.47$ & $-0.53$ & $6.408$ & $1.182$ & $0.003$ \\
\bottomrule
\end{tabular}
\vspace{-0.2cm}
\label{tab:nn_vs_phys_6}
\end{table*}

\endgroup

\section{Conclusions and Future Work}

A data-efficient PIML framework was proposed to model synchro-waveform dynamics of grid-integrated IBRs. The method leverages time-synchronized WMU measurements at both the IBR terminal and the grid side to embed circuit equations directly into the learning process. This enables accurate data-driven modeling using substantially fewer instances of disturbance data than purely data-driven methods. Across multiple sampling rates, the proposed method achieved lower estimation error and  robust performance.

The PIML framework also supports joint estimation of unknown circuit parameters together with the IBR transient response. This feature is particularly relevant in practical settings where line parameters are uncertain or unavailable. 

Future work will consider more complex network models, including radial and meshed topologies, measurement noise, multiple IBRs, and three-phase unbalanced conditions.



\bibliographystyle{IEEEtran}
\bibliography{Ref}

@misc{NERC_Report_2,
    AUTHOR = {{North American Electric Reliability Corporation}},
     TITLE = {{Multiple Solar PV Disturbances in CAISO}},
	HOWPUBLISHED = {NERC Report },
     year = 2022,
}

@ARTICLE{Hossein_letter,
  author={Mohsenzadeh-Yazdi, Hossein and Li, Chester and Mohsenian-Rad, Hamed},
  journal={IEEE Trans. on Smart Grid}, 
  title={{ IBR Responses During a Real-World System-Wide Disturbance: Synchro-Waveform Data Analysis, Pattern Classification, and Engineering Implications}}, 
  year={2025},
  volume={},
  number={},
  pages={1-1},
  doi={10.1109/TSG.2025.3564102}
}

@ARTICLE{subcycle_overvoltage,
  author={Fan, Lingling and Miao, Zhixin and Zhang, Miao},
  journal={IEEE Trans. on Power Delivery}, 
  title={{Subcycle Overvoltage Dynamics in Solar PVs}}, 
  year={2021},
  volume={},
  pages={1847-1858}
}

@article{synchro_waveform_window,
 title={{Synchro-waveforms: A Window to the Future of Power Systems Data Analytics}},
  author={H. {Mohsenian-Rad}  and W. {Xu}},
  journal={IEEE Power and Energy Magazine},
  volume={21},
  number={5},
  pages={68-77},
  year={2023},
}

@article{synchro_waveform_mladen,
 title={{Synchro-Waveforms in Wide-Area Monitoring, Control, and Protection: Real-World Examples and Future Opportunities}},
  author={H. {Mohsenian-Rad}  and M.~Kezunovic and F.~ Rahmatian},
  journal={IEEE Power and Energy Magazine},
  volume={23},
  number={1},
  pages={69-80},
  year={2025},
}

@ARTICLE{Hossein-Journal,
  author={Mohsenzadeh-Yazdi, Hossein and Ahmadi-Gorjayi, Fatemeh and Mohsenian-Rad, Hamed},
  journal={IEEE Trans. on Power Delivery}, 
  title={{Data-Driven Modeling of Sub-Cycle Dynamics of Inverter-Based Resources Using Real-World Synchro-Waveform Measurements}}, 
  year={2025},
  volume={40},
  number={4},
  pages={2314-2326},
  doi={10.1109/TPWRD.2025.3576589}}

@article{cai2016modelling,
  title={{Modelling, analysis and control design of a two-stage photovoltaic generation system}},
  author={Cai, Hongda and Xiang, Ji and Wei, Wei},
  journal={IET Renewable Power Generation},
  volume={10},
  number={8},
  pages={1195--1203},
  year={2016},
}

@article{physics_1,
  title={{Dynamic modeling and performance analysis of a grid-connected current-source inverter-based photovoltaic system}},
  author={P. P. {Dash} and M. {kazerani}},
  journal={IEEE Trans. on Sustainable Energy},
  volume={},
  no={4},
  pages={443--450},
  year={2011},
  publisher={IEEE}
}

@article{physics_2,
  title={{Dynamics of inverter-based resources in weak distribution grids}},
  author={M. {Ghazavi Dozein} and B. C. {Pal} and P. {Mancarella}},
  journal={IEEE Trans. on Power Systems},
  volume={37},
  no={5},
  pages={3682-3692},
  year={2022},
  publisher={IEEE}
}

@INPROCEEDINGS{palberz,
author={P. {Khaledian} and A. {Shahsavari} and H. {Mohsenian-Rad}}, 
booktitle={Proc. of the IEEE PES ISGT}, 
title={{Event-Based Dynamic Response Modeling of Large Behind-the-Meter Solar Farms: A Data-Driven Method Based on Real-World Data}}, 
address = {Washington, DC},
year={2023},
}

@article{Fatemeh_letter,
 title={{Data-Driven Models for Sub-Cycle Dynamic Response of Inverter-Based Resources Using WMU Measurements}},
  author={F. {Ahmadi-Gorjayi} and H. M. {Rad}},
  journal={IEEE Trans. on Smart Grid},
  volume={},
    pages={4125--4128},
  year={2023},
}

@ARTICLE{OPF-PINN,
  author={Lopez-Garcia, Tania B. and Domínguez-Navarro, José Antonio},
  journal={IEEE Trans. on Power Systems}, 
  title={{Optimal Power Flow With Physics-Informed Typed Graph Neural Networks}}, 
  year={2025},
  volume={40},
  number={1},
  pages={381-393},
  doi={10.1109/TPWRS.2024.3394371}}

@ARTICLE{Data_Anomaly_PINN,
  author={Banerjee, P. and Sivaramakrishnan, V. and Srivastava, A. K.},
  journal={IEEE Trans. on Industry Applications}, 
  title={{Decentralized Modular Nonlinear Physics-Informed Neural Network (mnPINN) for Synchrophasor Data Anomaly Detection}}, 
  year={2025},
  volume={61},
  number={2},
  pages={2490-2503},
  doi={10.1109/TIA.2025.3529822}}

@ARTICLE{DSE_PINN,
  author={de Jongh, Steven and Mueller, Felicitas and Cañizares, Claudio A. and Leibfried, Thomas and Bhattacharya, Kankar},
  journal={IEEE Trans. on Smart Grid}, 
  title={{Distribution Grid State Estimation With Limited Actual and Pseudo Measurements}}, 
  year={2025},
  volume={16},
  number={5},
  pages={3638-3652},
  doi={10.1109/TSG.2025.3576697}}

@ARTICLE{RPSSE_PINN,
  author={Falas, Solon and Asprou, Markos and Konstantinou, Charalambos and Michael, Maria},
  journal={IEEE Trans. on Industrial Informatics}, 
  title={{Robust Power System State Estimation Using Physics-Informed Neural Networks}}, 
  year={2025},
  volume={21},
    pages={8057-8067},
  doi={10.1109/TII.2025.3582293}}

@ARTICLE{Line_par_PINN,
  author={Wang, Wenyu and Yu, Nanpeng},
  journal={IEEE Trans. on Power Systems}, 
  title={{Estimate Three-Phase Distribution Line Parameters With Physics-Informed Graphical Learning Method}}, 
  year={2022},
  volume={37},
  number={5},
  pages={3577-3591},
  doi={10.1109/TPWRS.2021.3134952}}

@ARTICLE{MR-PP-GEK:19,
  author={Raissi, Maziar and Perdikaris, Pa and Karniadakis, George E.},
  journal={Journal of Computational Physics}, 
  title={{Physics-informed neural networks: A deep learning framework for solving forward and inverse problems involving nonlinear partial differential equations}}, 
  year={2019},
  volume={378},
  number={},
  pages={686-707},
  }

@ARTICLE{GSM-AV-SC:20,
  author={Misyris, Ga S. and Venzke, Aa and Chatzivasileiadis, Sa.},
  journal={Proceedings of the IEEE Power \& Energy Society General Meeting}, 
  title={{Physics-Informed Neural Networks for Power Systems}}, 
  year={2020},
  volume={92},
  number={3},
  pages={88},
  }

@book{Hamed_book,
  title={Smart Grid Sensors: Principles and Applications},
  author={Hamed Mohsenian-Rad},
  year={2022},
  month = {Apr.},
  publisher={Cambridge University Press, UK}
}

\end{document}